\documentclass{elsart3p}

\usepackage{graphics}
\usepackage{graphicx}
\usepackage{amssymb}

\begin{document}

\begin{frontmatter}

\title{Site-selective $^{63}$Cu NMR study of the vortex cores of Tl$_{2}$Ba$_{2}$CuO$_{6+\delta}$}

\author[label1]{Y. Itoh}
\ead{itoh@kuchem.kyoto-u.ac.jp}
\corauth{Tel: +81-75-753-3989.  Fax: +81-75-753-4000}
\author[label1]{C. Michioka} 
\author[label1]{K. Yoshimura}
\author[label2]{A. Hayashi\thanksref{thank1}} 
\author[label2]{Y. Ueda}
\address[label1]{Department of Chemistry, Graduate School of Science, Kyoto University, Kyoto 606-8502, Japan}
\address[label2]{Institute for Solid State Physics, University of Tokyo, 5-1-15 Kashiwanoha, Kashiwa, Chiba 277-8581, Japan}

\thanks[thank1]{The present author moved from ISSP.}

\begin{abstract}
We report site-selective $^{63}$Cu NMR studies of the vortex core states
of an overdoped Tl$_{2}$Ba$_{2}$CuO$_{6+\delta}$ with $T_{c}$ = 85 K. 
We observed a relatively high density of low-energy quasi-particle excitations at the vortex cores in a magnetic field o¶f 7.4847 T along the $c$ axis, in contrast to YBa$_{2}$Cu$_{3}$O$_{7-\delta}$.  
\end{abstract}

\end{frontmatter}

\section{Introduction}
The vortex core magnetism of high-$T_{c}$ cuprate superconductors has attracted great interests. This is the first report of site-selective $^{63}$Cu NMR study 
for the vortex core states of an overdoped Tl$_{2}$Ba$_{2}$CuO$_{6+\delta}$ (TL2201) with $T_\mathrm{c}$ = 85 K. 
Although a quadrupole Cu nuclear is coupled by an electric field gradient in TL2201, 
we noticed that the central transition line ($I_{z}$ = 1/2 $\leftrightarrow$ $-$1/2) of the  Cu (spin $I$ = 3/2) NMR at a magnetic field $H$ along the maximal principal axis ($c$ axis) of the electric field gradient tensor is purely magnetic.       
As to YBa$_{2}$Cu$_{3}$O$_{7-\delta}$ (Y1237) with $T_\mathrm{c}$ = 92 K,  both $^{17}$O and $^{63}$Cu nuclear spin-lattice relaxation rates 1/$T_{1}$ inside the vortex cores are reported to be enhanced more largely than those outside the cores only below about 20 K \cite{Curro,Mitrovic,ZPNQR}. 
For TL2201, we observed that the $^{63}$Cu NMR 1/$T_{1}$ near the vortex cores is enhanced 
more largely than that away from the cores just at $T_\mathrm{c}$ and shows a Korringa-like behavior from $T_{c}$ to 10 K.   

\section{Experiments}
A powder sample was magnetically aligned along the $c$ axis and was characterized more than ten years ago \cite{Kambe} and recently \cite{ItohTL2201}. 

\begin{figure}[ht]
\begin{center}
\includegraphics[width=0.50\textwidth]{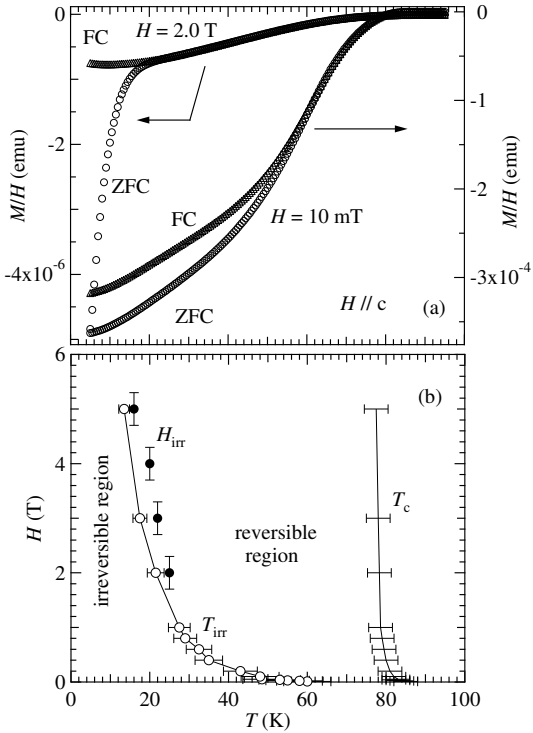}
\end{center}
\caption{(a) Magnetic susceptibility $M$/$H$ after cooled at a zero magnetic field (ZFC) and at finite magnetic fields
of $H$ = 10 mT and 2 T (FC) along the $c$ axis. 
(b) A magnetic phase diagram on an irreversible line along the $c$ axis. 
Open circles are the irreversible temperatures at the fixed fields. 
Closed circles are the irreversible fields at the fixed temperatures.}
\end{figure}

\begin{figure}[ht]
\begin{center}
\includegraphics[width=0.55\textwidth]{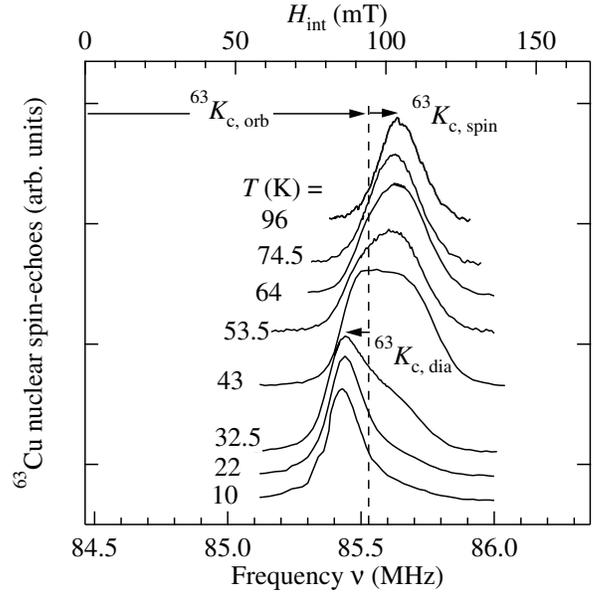}
\end{center}
\caption{Central transition lines  ($I_{z}$ = 1/2 $\leftrightarrow$ $-$1/2) of $^{63}$Cu NMR frequency spectra with cooling at 7.4847 T along the $c$ axis. The reference frequency $\nu_{0}$ = 84.465 MHz is given by $\nu_{0}$=$^{63}\gamma_{n}H$ with the $^{63}$Cu nuclear gyromagnetic ratio $^{63}\gamma_{n}/2\pi$ = 11.285 MHz/T. The top axis is an internal magnetic field $H_\mathrm{int}=(\nu-\nu_{0})/^{63}\gamma_{n}$.}
\end{figure}

\begin{figure}[ht]
\begin{center}
\includegraphics[width=0.50\textwidth]{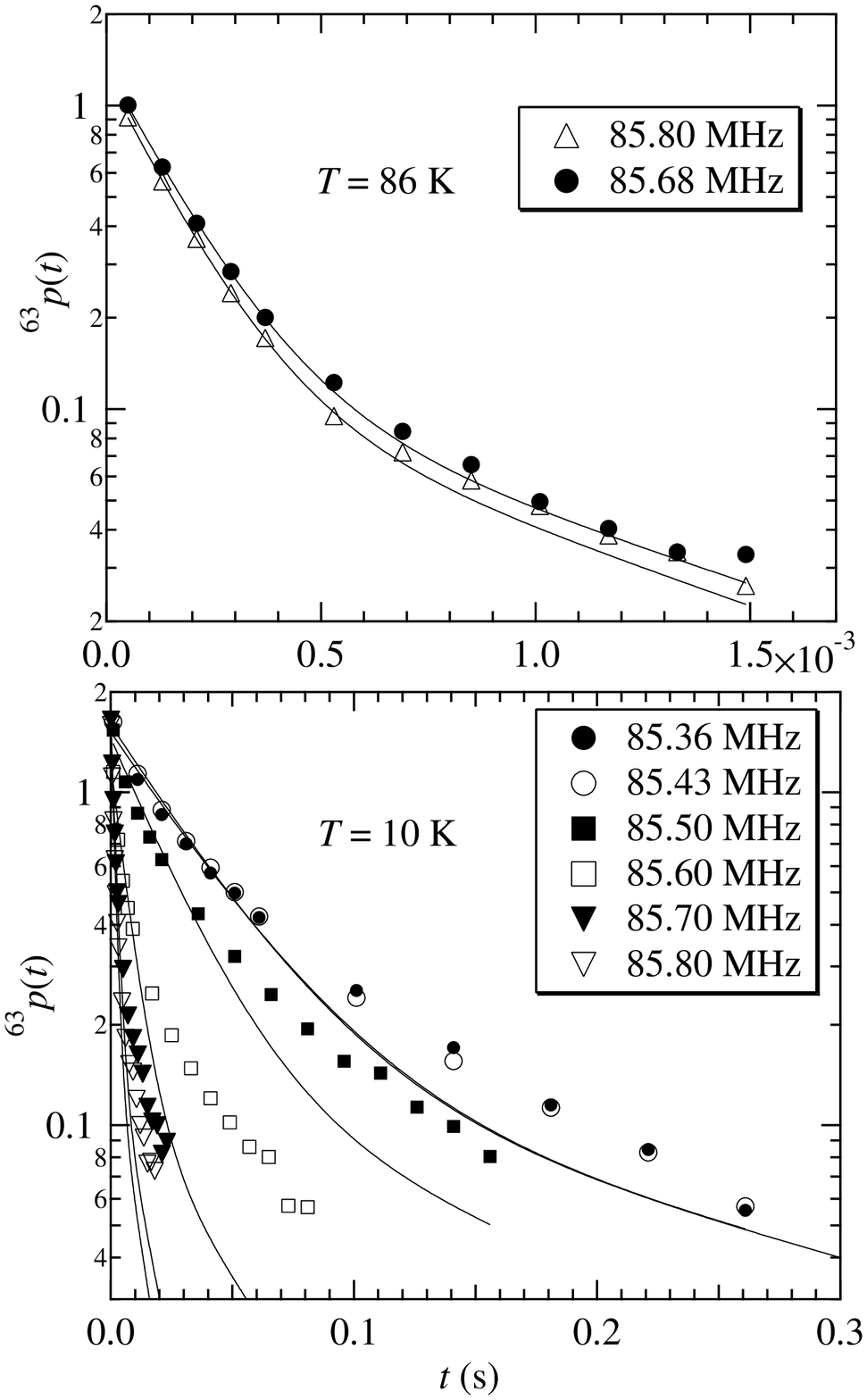}
\end{center}
\caption{$^{63}$Cu nuclear spin-echo recovery curves $^{63}p(t)$ 
of the central transition line ($I_{z}$ = 1/2 $\leftrightarrow$ $-$1/2) of the  $^{63}$Cu NMR 
as functions of frequency at $T$ = 86 K (upper) and 10 K (lower).
The solid curves are the least-squares fits by theoretical functions (see the text).}
\end{figure}

Magnetization $M$ was measured by a superconducting quantum interference device (SQUID) magnetometer. 
The irreversible temperature $T_\mathrm{irr}$ was defined as a bifurcation temperature of $M$/$H$ after cooled at a zero magnetic field (ZFC)
and at finite magnetic fields of $H$ = 10 mT $-$ 5.0 T (FC). 
Typical magnetic susceptibility $M$/$H$ at 10 mT and 2.0 T is shown in Fig. 1(a). 
The $T_\mathrm{irr}$ was found to be quickly suppressed as the magnetic field $H$ was increased. 
The irreversible line of TL2201 in Fig. 1(b) is similar to that of Bi$_2$Sr$_2$CaCu$_2$O$_8$ (Bi2212). 
The anomaly observed in Tl NMR at 20 K and at 2.0 T is associated 
with a vortex freezing effect  across the irreversible line \cite{ItohTL2201}. 

Site-selective $^{63}$Cu NMR experiments were performed by  a phase-coherent-type pulsed spectrometer while cooling in a magnetic field of $H$ = 7.4847 T. 
The $^{63}$Cu NMR frequency spectra 
were measured with quadrature detection.
The nuclear spin-echoes were recorded
as functions of frequency $\nu$ 
while $\nu$ was changed point by point.  
The $^{63}$Cu nuclear spin-lattice relaxation curves
$^{63}p(t)\equiv 1-M(t)/M(\infty)$ (recovery curves) 
of the nuclear spin-echo amplitude $M(t)$
were measured by an inversion recovery technique,
as functions of time $t$ after an inversion pulse.  

\section{$^{63}$Cu NMR Results and Discussion}
Fig. 2 shows temperature dependence of the central transition line ($I_{z}$ = 1/2 $\leftrightarrow$ $-$1/2) of the  $^{63}$Cu NMR while cooling at $H$ = 7.4847 T along the $c$ axis below 96 K. 
In the normal state, the positive Knight shift $K_\mathrm{c}$ of the observed sharp line is 
the sum of an orbital shift $K_\mathrm{c, orb}$ and a spin shift $K_\mathrm{c, spin}$ \cite{Kambe}. Below $T_\mathrm{c}$ down to 22 K, the $^{63}$Cu NMR spectra are composed of two spectra with a finite spin shift and a reduced Knight shift $K_\mathrm{c} < K_\mathrm{c, orb}$.
The reduced shift is due to a superconducting diamagnetic shift $K_\mathrm{c, dia}$
in a vortex lattice. 
These spectra indicate the coexistence of vortex solid and liquid, 
similarly to the $^{205}$Tl NMR spectra at 2.0 T \cite{ItohTL2201}.  
Below 22 K, the Redfield pattern indicates the vortex lattice. 
In contrast to Bi2212 \cite{Bi2212} and HgBa$_2$CuO$_{4+\delta}$ \cite{Hg1201}, the motional narrowing effect on the Cu NMR linewidth was not observed in TL2201
at 7.4847 T.  

\begin{figure}[ht]
\begin{center}
\includegraphics[width=0.55\textwidth]{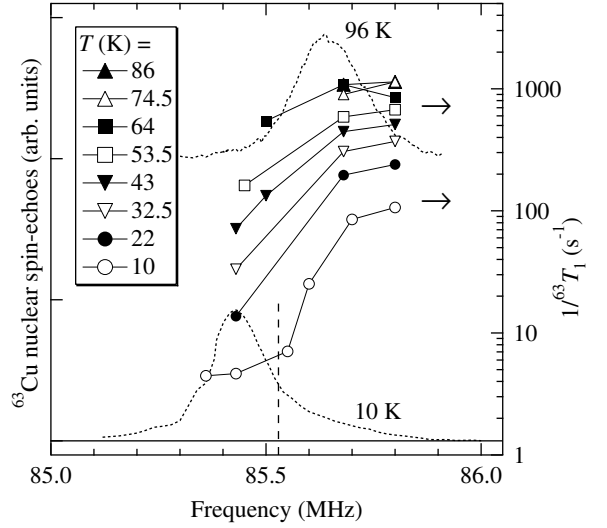}
\end{center}
\caption{Temperature dependence of frequency distribution of $^{63}$Cu nuclear spin-lattice relaxation rate 1/$^{63}T_{1}$ (the right axis).}
\end{figure}

\begin{figure}[ht]
\begin{center}
\includegraphics[width=0.50\textwidth]{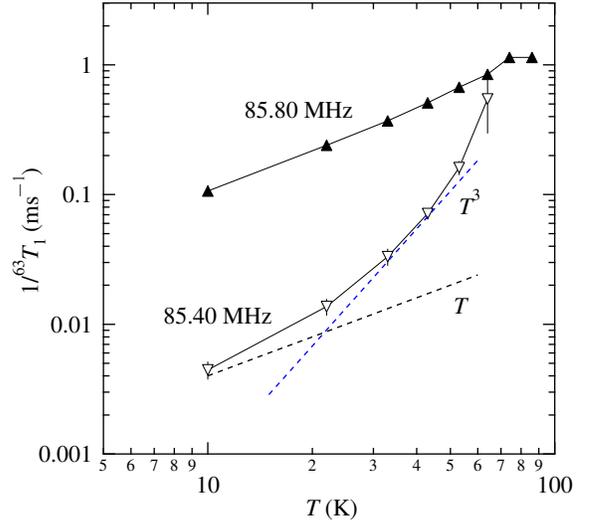}
\end{center}
\caption{Log-log plots of $^{63}$Cu nuclear spin-lattice relaxation rate 1/$^{63}T_{1}$ as a function of temperature at 85.68 MHz around the vortex cores and at 85.40 MHz away from the vortex cores.}
\end{figure}

\begin{figure}[ht]
\begin{center}
\includegraphics[width=0.50\textwidth]{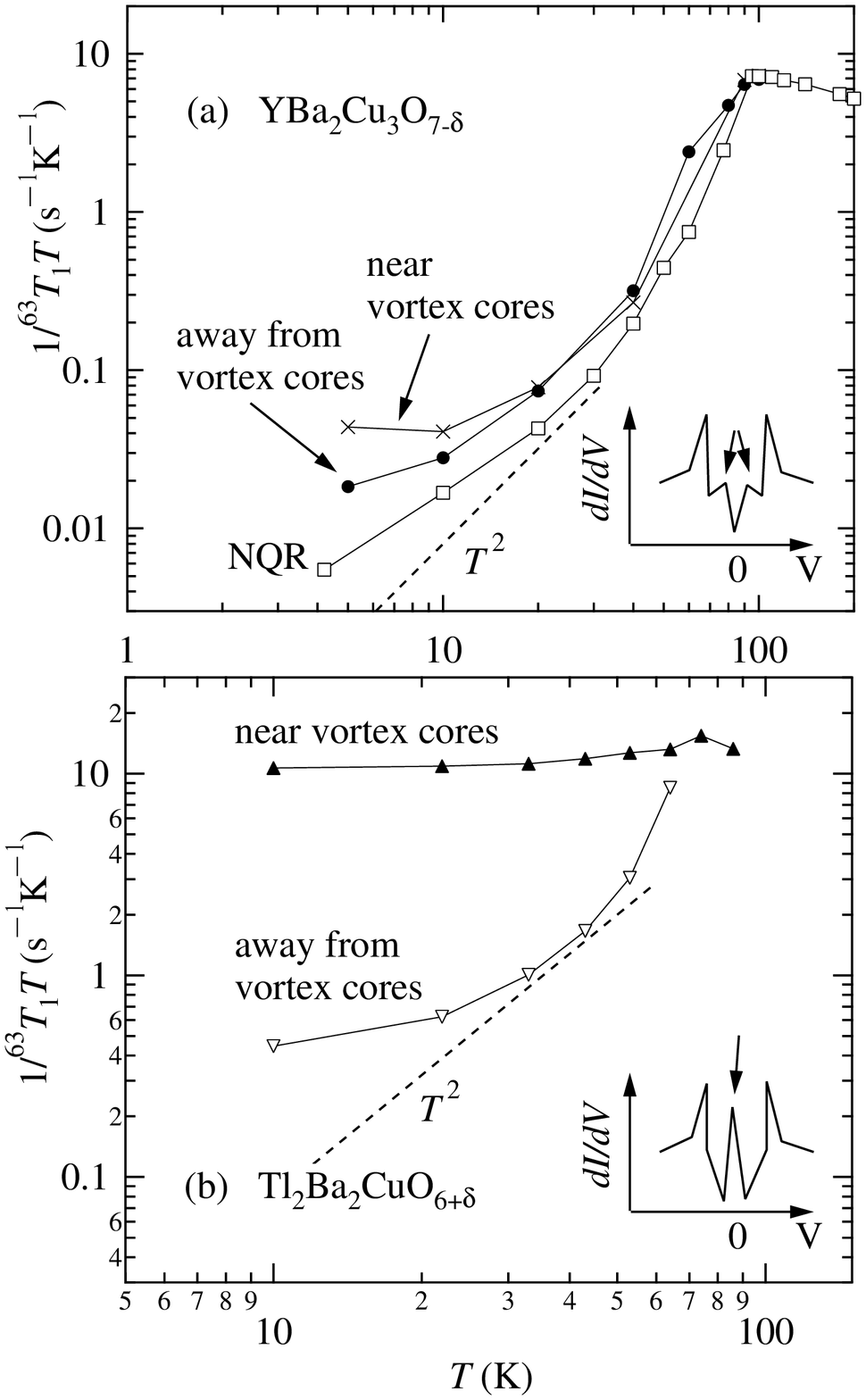}
\end{center}
\caption{Log-log plots of $^{63}$Cu nuclear spin-lattice relaxation rate divided by temperature $T$, 1/$^{63}T_{1}T$,  for Y1237 (a) reproduced from Ref. \cite{ZPNQR} and for TL2201 (b).
Each inset figure illustrates a conductance spectrum d$I$/d$V$ near the vortex cores.
The zero-bias conductance spectrum splits near the cores in Y1237 \cite{STS}, 
while it must show a peak in TL2201. }
\end{figure}

Fig. 3 shows the frequency distribution of the $^{63}$Cu NMR recovery curves $^{63}p(t)$
at 86 and 10 K. 
In the normal state at 86 K, a slight frequency distribution was observed, similarly to $^{17}$O NMR for YBa$_2$Cu$_4$O$_8$  \cite{Kumagai}, maybe due to slight misorientation of the powder grains. 
At 10 K, the relaxation is slower away from the vortex cores.   
The solid curves are the least-squares fits by theoretical functions
of $p(t)$=$p$(0)[0.1exp($-t$/$T_{1}$)+0.9exp($-$6$t$/$T_{1}$)]
for a magnetic transition of $I_{z}$ = 1/2 $\leftrightarrow$ $-$1/2. 
Although the fits were not so satisfactory at 10 K, the relaxation rates 1/$^{63}T_{1}$ were tentatively estimated from these fits. 

Fig. 4 shows temperature dependence of frequency distribution of  the estimated 1/$^{63}T_{1}$. 
With cooling down, a larger frequency distribution  of 1/$^{63}T_{1}$ was observed. 
In contrast to the $^{205}$Tl NMR \cite{ItohTL2201}, no clear effect of the vortex freezing was  observed.   

Fig. 5 shows temperature dependence of 1/$^{63}T_{1}$ at 85.68 MHz near the vortex cores and at 85.40 MHz away from the cores. 
At $T_\mathrm{c}$, 1/$^{63}T_{1}$ away from the cores quickly decreases in a function of  $T^3$. 
Below 20 K, it approaches a $T$-linear function. 
From $T_\mathrm{c}$ to 10 K, 1/$^{63}T_{1}$ near the cores shows a Korringa-like $T$-linear behavior. 
These results are different from those for Y1237 \cite{Curro,Mitrovic,ZPNQR} 
but similar to the theoretical 1/$T_{1}$ due to the spatial distribution of a local density of electron states \cite{TIM} and also to the Zn-induced effect on Y1237 \cite{Y1237Zn}.
The theory does not include antiferromagnetic correlation. 
Although 1/$^{63}T_{1}$ between $T_\mathrm{c}$ and 43 K might be affected by
a direct process of overdamped motion of pancake vortices \cite{pancake},
the Korringa-like 1/$^{63}T_{1}$ in a vortex lattice below 32.5 K indicates 
a relatively high density of normal quasi-particle excitations inside the vortex cores in TL2201. 

Fig. 6 shows the $^{63}$Cu nuclear spin-lattice relaxation rate divided by temperature, 1/$^{63}T_{1}T$,  for Y1237 reproduced from Ref. \cite{ZPNQR} and for TL2201. 
The difference in  1/$^{63}T_{1}T$ inside and outside the vortex cores is larger in TL2201 than in Y1237. 

Scanning tunneling spectroscopy (STS) studies for Y1237 indicate the split of a zero-bias peak in the conductance spectra near the vortex cores at a magnetic field \cite{STS}. 
The impurity Zn-substitution effects are known to induce the zero-bias conductance peak in the STS spectra near the Zn impurity \cite{Pan} and 
the difference in 1/$^{63}T_{1}T$ near and away from the Zn just below $T_\mathrm{c}$ \cite{Y1237Zn}.     
Thus, the STS conductance spectrum for TL2201 might indicate the zero-bias conductance peak near the vortex cores, which is illustrated in the inset of Fig. 6(b). 
The nature of Andreev bound states at the vortex cores, {\it e.g.},
the degree of the split of the zero-bias conductance peak near the vortex cores
might depend on the degree of enhancement of underlying antiferromagnetic correlation. 

\section{Conclusion}
We report for the first time the site-selective $^{63}$Cu NMR studies of the vortex core magnetism
for TL2201 with $T_\mathrm{c}$ = 85 K at about 7.5 T along the $c$ axis. 
The difference in $^{63}$Cu nuclear spin-lattice relaxation rate 1/$^{63}T_{1}$ inside and outside the vortex cores was observed in TL2201 just below $T_\mathrm{c}$ = 85 K, 
in contrast to that in 1/$^{63}T_{1}$ in Y1237 below about 20 K. 
The Korringa-like behavior of 1/$^{63}T_{1}$ near the vortex cores indicates a high density of the normal quasi-particle excitations inside the vortex cores in TL2201. 

\begin{ack}
We thank Professor N. Nishida for fruitful discussion. 
This study was supported by a Grant-in-Aid 
for Science Research on Priority Area,
``Invention of anomalous quantum materials" from the Ministry of 
Education, Science, Sports and Culture of Japan (Grant No. 16076210).
\end{ack}


\end{document}